\documentclass[useAMS,a4paper]{mn2e}
\input{psfig}
\usepackage{amsfonts,amssymb,amsmath}
\usepackage{aas_macros}
\usepackage{times,varioref,color,multirow,textcomp}
\usepackage[
    dvips,
    a4paper=true,
    plainpages=true,
    pdfpagelabels,
    bookmarks=true,
    bookmarksopen=false,
    bookmarksopenlevel=2
    bookmarksnumbered=true,
    bookmarkstype=toc,
    colorlinks=true,
    citecolor=red,
    linkcolor=blue,
    menucolor=green,
    urlcolor=magenta,
]{hyperref}
  
\begin{document}


\title[stellar mass gap in galaxy groups]
      {The gap of stellar mass in galaxy groups: another perspective of the Too-big-To-Fail problem in the Milky Way}
\author[Kang, Wang \& Luo]
       {X. Kang$^{1}$, L. Wang$^{1}$ \& Y. Luo$^{1}$\thanks{E-mail:kangxi@pmo.ac.cn}\\
        $^1$Purple Mountain Observatory, the Partner Group of MPI f\"ur Astronomie, 2 West Beijing Road, Nanjing 210008, China}

\date{}

\pagerange{\pageref{firstpage}--\pageref{lastpage}}
\pubyear{2015}

\maketitle

\label{firstpage}


\begin{abstract}

The Milky Way  presents the too-big-to-fail (TBTF)  problem that there
are  two observed  satellite galaxies  with maximum  circular velocity
larger than 55km/s, and others have velocity less than 25km/s, but the
cold dark matter model predicts there should be more than 10 subhaloes
with  velocity  larger  than  25km/s.  Those  massive  subhaloes  with
$25km/s<V_{max}<55km/s$ should not have failed  to form stars.  The TBTF
problem severely challenges  the CDM model.  Most  efforts are seeking
the effects  of baryonic  feedback, decreasing the  mass of  the Milky
Way,  changing the  properties of  dark matter,  so as  to assign  the
observed low-velocity  satellites into the massive  subhaloes found in
simulations. However, the  TBTF problem can be avoided if  the MW have
not accreted  subhaloes with velocity  between $25km/s<V_{max}<55km/s$
although the probability of such a gap  is lower as $\sim 1\%$ and can
not be tested  against observations. In this work we  study the gap in
stellar mass of satellite galaxies  using the SDSS group catalogue and
a semi-analytical model. We find that there are 1-2\% of galaxy groups
with a large gap in the  stellar mass of their satellites.  These 'big
gap' groups  have accreted less  massive subhaloes in  their formation
history  and   naturally  display   a  gap  between   their  satellite
galaxies.  If   extrapolating  our  results   to  the  Milky   Way  is
appropriate, we conclude that it is very likely that our Milky Way has
not  accreted  enough massive  subhaloes  to  host those  low-velocity
satellites, and the TBTF problem is naturally avoided.

\end{abstract}



\section{Introduction}
\label{sec:intro}

In the last two decades  there are three prominent problems concerning
the  satellite galaxies  in  the Milky  Way (MW).   The  first is  the
'missing satellite problem' that cold dark matter (CDM) model predicts
hundred of subhaloes but only a dozen classical satellite galaxies are
observed (Klypin  et al. 1999;  Moore et  al. 1999).  With  more faint
satellites being found  and development of more  realistic modeling of
galaxy formation,  the tension in  this problem is  greatly alleviated
(e.g.,  Gendin  2000; Benson  et  al.   2002;  Koposov et  al.   2008;
Macci{\`o} et  al. 2010; Font et  al.  2011).  The second  is the thin
planar distribution of the classical  satellites (Kroupa et al. 2005).
Analysis using  simulation found that  the probability of such  a thin
disk  ranges from  a few  percent to  30\% (e.g.,  Kang et  al.  2005;
Zentner et  al.  2005; Libeskind et  al. 2009; Wang et  al. 2013), but
the chance is  much lower considering the co-rotation  of the majority
of the satellites in the plane  (e.g., Pawlowski \& Kroupa 2013).  The
third  is   the  too-big-to-fail  problem  (TBTF;   Boylan-Kolchin  et
al. 2011; 2012) that only two  (LMC, SMC) of the dozen satellites have
$V_{max}>55km/s$ and  others have  $V_{max}<25km/s$, but  CDM predicts
more than  10 subhaloes  with $V_{max}>25km/s$  in the  Milky Way-size
halo and they  should not have failed to form  stars. The TBTF problem
is also seen  in the local group (Garrison-Kimmel et  al. 2014; Sawala
et al. 2015; Brooks \& Di Cintio 2015).

The  TBTF  problem  severely   challenges  the  CDM  model.  The  most
straightforward solution is to  lower the circular velocity or central
density of  the massive subhaloes in  simulations so as  to host those
observed satellite  galaxies. Most efforts are invoked  to include the
baryonic  feedback  to decrease  the  central  density  of subhalo  in
simulations  (e.g., Governato  et  al.  2012;  Zolotov  et al.   2012;
Brooks  et   al.   2013;  Brooks   \&  Zolotov  2014;  Aaron   et  al.
2016).  However,  the  baryonic   effects  are  still  debated  (e.g.,
Boylan-Kolchin et  al.  2012; Garrison-Kimmel et  al.  2013; Pawlowski
et  al.  2015).  Alternative  solution to  lower  the subhalo  central
density is  to change the dark  matter property, such as  using a warm
dark matter (e.g.,  Lovell et al.  2012; Maccio et  al.  2013), or the
self-interacting dark matter (e.g., Vogelsberger et al. 2012; Rocha et
al. 2013).

On the other hand, the TBTF problem can be alleviated if there is only
a few massive subhaloes in the MW  mass halo.  Wang et al. (2012) have
shown  that  the number  of  subhalo  with $V_{max}>25km/s$  decreases
quickly with  halo mass,  and if  requiring the MW  has at  most three
subhaloes  with $V_{max}>25km/s$,  the halo  mass of  the MW  would be
lower than $1.4\times  10^{12}M_{\odot}$. However, a lower  MW mass is
difficult to reconcile with the occurrence of the two observed massive
satellites (LMC,  SMC) with  $V_{max}>55km/s$.  Boylan-Kolchin  et al.
(2010) have shown that a MW  halo with mass of $10^{12}M_{\odot}$ will
have less  than 10\% chance  to host two  satellites as bright  as the
Magellanic  Clouds.  Such  a  lower  chance  is  also  observationally
supported (Liu  et al.  2011). Thus the  occurrence of  the Magellanic
Clouds   and   avoidance   of   more   than   three   subhaloes   with
$V_{max}>25km/s$ together can put a  strong constraints on the MW mass
(Cautun et al. 2014).

A more accurate description of  the MW satellite velocity distribution
is that  there are no satellite  galaxies with $25km/s<V_{max}<55km/s$
(e.g., Jiang \&  van den Bosch 2016). Cautun et  al. (2014) found that
the probability of having such a wide gap in velocity space is about 1\%
in the CDM model.  Jiang \&  van den Bosch (2016) find the probability
is even  lower as 0.1\%  using a  monte-carlo method.  If  the subhalo
population of  the MW  happens to  has such  a gap or distribution,  the TBTF
problem  is naturally  avoided. However,  due  to lack  of large  data
sample analog  to the  MW mass/luminosity  (along with  measurement of
their  satellites  velocity)  from observations,  such  a  statistical
probability can not be tested.

To find the occurrence of systems alike the MW in terms of a big gap in
the distribution of its satellites,  one need large sample of galaxies
with well determined satellite galaxy population.  Using galaxies from
the Sloan Digital Sky Survey (SDSS), Yang et al.  (2012) constructed a
large  sample   of  galaxy  groups  with  member   galaxies  are  well
determined.  However,  the SDSS lacks  velocity dispersion measurement
of the main galaxy sample, we  thus describe the gap using the stellar
mass of satellite galaxies in the group.  We ask one related question:
are there any  galaxy groups which display similar  gap in the stellar
mass of  their satellites as in the  MW?  We term those  groups with a
big  gap  in  the  stellar  mass  of their  satellites  as  'big  gap'
groups.\footnote{Our  'big  gap'  definition  is  different  from  the
  traditional  definition  of  'fossil   group'  which  refer  to  the
  luminosity/stellar mass gap between the central and the most massive
  satellite galaxy (e.g., Ponman et  al. 1994), while we are referring
  to  the gap  between  the satellite  galaxies  themselves.}  We  first
identify 'big gap' groups from  the Yang et al. (2012) group catalogue
and compare their properties to the predictions from a semi-analytical
model (Kang et  al.  2012).  For those 'big gap'  groups in our model,
we look into their formation history and investigate the origin of the
gap in the stellar mass of their satellites.  We believe that although
we are  looking at  massive counterparts of  the MW, the  formation of
those 'big gap' groups can also shed light on the formation of the MW,
even on the nature of the TBTF problem.

The paper is organized as:  in Section.2 we briefly introduce the used
group  catalogue, model  galaxies and  how we  identify the  'big gap'
groups. In section.3 we compare  the data to the model predictions and
use the model galaxies to identify the origin of the 'big gap' groups.
We finally summarize and discuss in Section.4.

\section{Group catalogue and model galaxies}
\label{sec:methods}

We use the group catalogue constructed by Yang et al. (2012) which is
now                                                            publicly
available\footnote{http://gax.shao.ac.cn/data/Group.html}.  The  group
catalogue is based on the data from the SDSS Data Release 7 (Abazajian
et  al. 2009)  which contains  both photometric  and  spectroscopy of 
about 1 million galaxies  with Petrosian magnitude $r<17.77$. For each
galaxy  the stellar  mass  is estimated  using  the model  of Bell  et
al. (2003).  Using the stellar mass  and position of  each galaxy, the
group  catalogue is constructed  using the  halo-based group  finder of
Yang et al. (2005). For detail of constructing the group catalogue, we
refer  the reader  to Yang  et al.  (2012). For  each group,  the most
massive  member   galaxy  is called as the  central  galaxy and  others  are
satellites. In our work, we  select group with virial mass ($M_{vir}$)
larger  than  $10^{13}M_{\odot}$  and  with redshift  $z<0.1$.  Groups
selected in the  way is more robust and  contains more member galaxies
(on average with 8 satellite galaxies per group), and we have 9610
groups in total.

The model galaxies used are from  the semi-analytical model of Kang et
al.  (2012)  combined with a  cosmological N-body simulation  with the
WMAP seventh-year  cosmological parameters (Komatsu et  al. 2011). The
simulation  is  run  using  the Gadget-2  code  (Springel  2005)  with
$1024^{3}$ dark matter particles in a cube of $200Mpc/h$ on each side.
The semi-analytical  model includes key physics  governing galaxy
formation,  such as  gas cooling,  star formation,  supernova and  AGN
feedback. By grafting the model onto  the merger trees from the N-body
simulation, the model provides good match to the observed stellar mass
function  of  SDSS  and  the  galaxy  two-point  correlation  function
simultaneously  (Kang 2014).   Also the  model fits  the stellar  mass
functions of  satellite galaxies  in different  halo mass  obtained by
Yang et al. (2012) using the SDSS DR7. The success of  the model lay down
the basis for comparison with the data in this work.

As  said  before  the  MW  displays a  gap  between  the  two  massive
satellites  (LMC, SMC,  $V_{max}>55km/s$)  and the  third massive  one
Sagittarius  ($V_{max} \sim  25km/s$, see  Tab.1 in  Jiang \&  van den
Bosch 2016)  and references therein).   In the  SDSS DR7, there  is no
velocity  dispersion measurement  for the  main galaxies,  we have  to
translate this velocity  gap into the stellar mass  gap.  The observed
gap  in stellar  mass  between SMC  ($4.6\times 10^{8}M_{\odot}$)  and
Sagittarius ($2.1\times 10^{7}M_{\odot}$, see  Tab.4 in McConnachie et
at.   2012 for  compilation of  local  dwarfs) is  about $\Delta  \log
M_{\ast} \sim 1.3$.  Is this gap in stellar mass  expected for the two
satellites  based on  their  $V_{max}$ or is there any significant contribution from  the
stochastic  star formation  in them?   Rodriguez-Puebla et  al. (2013)
have shown  that the SMC,  Sagittarius well stand on  the extrapolated
stellar mass-$V_{max}$ relation (stellar  Tully-Fisher relation, Avila-Reese et
al. 2008) with a slope of $\sim 0.3$. The expected gap in stellar mass
between  the two  satellites from  the $M_{\ast}-V_{max}$  relation is
thus       $\Delta        \log       M_{\ast}        \sim       \Delta
\log(V_{max,SMC}/V_{max,Sag})/0.3 \simeq  1.1$, which is close  to the
observed one. It indicates that the stochasticity of star formation in
SMC and Sagittatius  is equal or not important, so  the gap of $\Delta
\lg M_{\ast} \sim  1$ is more physically related to  the difference in
mass/$V_{max}$ of the two satellites.

For the following analysis, we describe the gap in stellar mass between the 
satellite galaxies in a group as $\Delta_{ij} = \log M_{\ast,i} - \log
M_{\ast,j}$, where  $M_{\ast,i}$ is the  stellar mass of  the $i^{th}$
massive satellite  galaxy.  The  gap in  the MW  is then  described as
$\Delta_{23} > 1$.  To be exactly analogous to the  MW, we should look
into groups  with $\Delta_{23} > 1$.  In our simulation, we  have 2831
groups  with  $M_{vir}>10^{13}M_{\odot}$, and  only  13  of them  have
$\Delta_{23}>1$. To increase the  sample for statistical significance,
here we use $\Delta_{12}$ and  label those groups with $\Delta_{12}>1$
as  'big  gap' groups.   In  this  way we  then  have  40 groups  with
$\Delta_{12}>1$  from  our simulation,  and  it  enable us  to  derive
reliable formation history of them and  study the origin of these 'big
gap' groups.

\begin{figure*}
\centerline{\psfig{figure=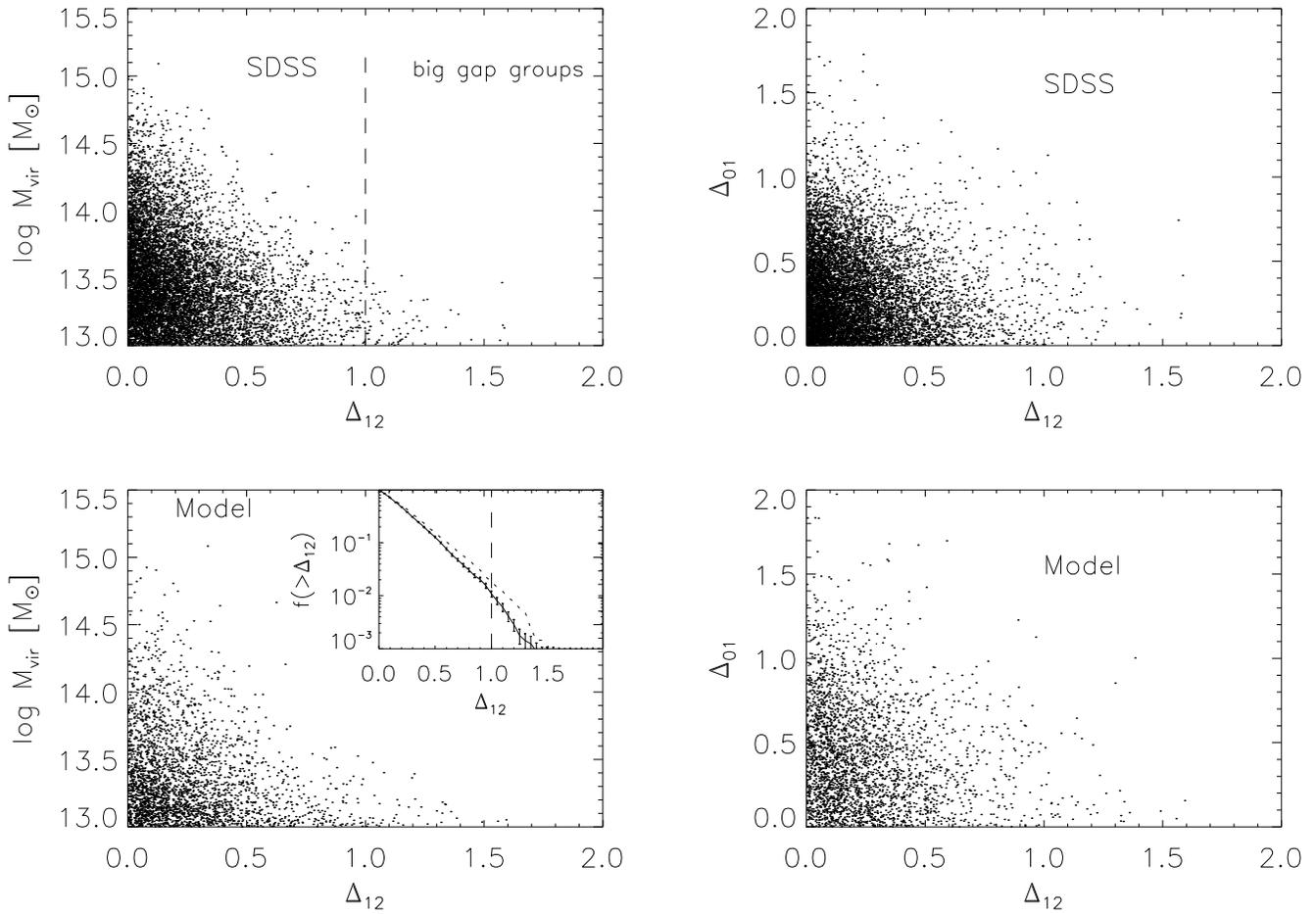,width=0.99\textwidth}}
\caption{The distribution of the gap in stellar mass between the first
  and second massive satellite galaxies in the group.  The left panels
  are the gap  as a function as  the halo mass, and the  right for the
  distribution   of   the   gap   $\Delta_{12}$   and   the   gap   in
  $\Delta_{01}$. The upper panels are results from the group catalogue
  (Yang et al.  2012) of the SDSS  DR7, and lower panels  are from the
  model. Inserted panel  in the lower left is  the cumulative fraction
  of groups as a function of $\Delta_{12}$.  The model predictions are
  very similar with the data.}
\label{fig:Mvir-Delta}
\end{figure*}

\section{Results}

In this  part we present the gap  distribution in the data  and in the
simulation,  and  check  if  the  model  galaxies  reproduce  the  gap
distribution seen in  the group catalogue from the SDSS.  Only in the case
the  model is able  to  describe  the  gap distribution  or  gap-halo
properties  correlation,  we  are   able  to  use  the simulated galaxies  to
investigate the origin of the 'big gap' groups.

Fig.\ref{fig:Mvir-Delta} shows  the scatter between  $\Delta_{12}$ and
the halo properties. The upper panels  are from the group catalogue of
Yang  et al.   (2012),  and  the lower  panels  are  results from  our
semi-analytical model.   The left panel show  the distribution between
$\Delta_{12}$ and  the virial mass of  the group. It is  seen that the
distribution from  the data  and the  model is  very similar:  the gap
distribution is dependent on group mass.  In massive groups the gap is
much narrow  and lower,  that the  difference in  stellar mass  of the
first,  second massive  satellite  galaxies is  smaller.  In  low-mass
groups, the distribution of $\Delta_{12}$ is  wider, with a tail up to
1.5. The vertical dashed line in the  upper left panel show the gap in
the MW,  and the groups to  the right of  the line are called  as 'big
gap' groups.   As the distribution  of $\Delta_{12}$ is a  function of
halo   mass,    we   select   a    narrow   mass   bin    with   $\log
M_{vir}=[13,13.5]M_{\odot}$, and  plot the cumulative  distribution of
$\Delta_{12}$ in these haloes in the  inserted panel in the left lower
panel, where the  solid line is for  the SDSS data and  dotted line is
our model.   It is seen that  the distribution is very  similar. There
are 1\%  'big gap' groups in  the data (solid line  with Poisson error
bar) and 2\% in the model.

The right  panels further test if  the gap distribution is  similar in
the data  and the model  by showing the scatter  between $\Delta_{12}$
and  $\Delta_{01}$, where  $\Delta_{01}$ is  the gap  in stellar  mass
between the  central galaxy and  the most massive  satellite.  Usually
$\Delta_{01}>1.0$  is  an indication  of  a  fossil group  (Ponman  et
al. 1994) and a fossil group is believed to has formed early and being
relaxed for  a longer  time, so most  massive satellite  galaxies have
merged with  the central galaxy (e.g.,  Jone et al. 2003;  D'Onghia et
al.   2005;    von   Benda-Beckmann   et   al.    2008;   Kundert   et
al. 2015). Firstly  seen is that the scatter  distribution between the
data and  the model is again  very similar, indicating the  model well
reproduce the properties of observed galaxies. It is also seen that in
either   fossil  groups   ($\Delta_{01}>1$)   or   'big  gap'   groups
($\Delta_{12}>1$),    the   distributions    of   $\Delta_{01}$    and
$\Delta_{12}$  are  not  strongly   correlated,  indicating  that  the
formation of fossil  groups and 'big gap' groups are  not related. The
formation of  a fossil group  is not due  to the selective  mergers of
satellite galaxies,  ie, those satellite  with mass between  the first
and second massive satellites.

\begin{figure}
\centerline{\psfig{figure=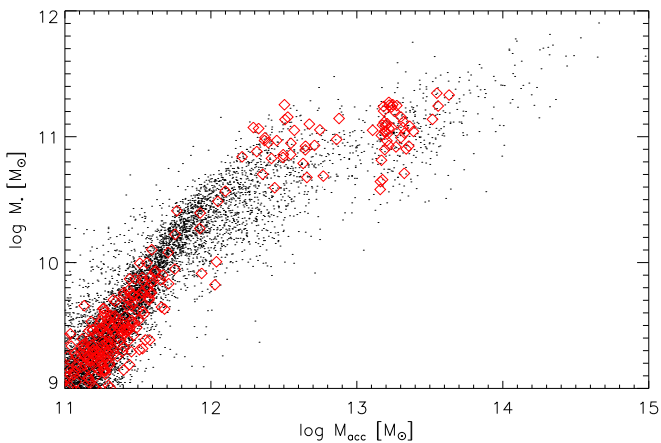,width=0.49\textwidth}}
\caption{The  relation between  the  stellar mass  and accretion  halo
  mass.  Black dots  are  for all  galaxies and  red  squares are  for
  galaxies in the  'big gap' groups (stellar mass differs  by a factor
  of  ten  between  the  most  and  second  massive  satellites).  The
  distribution of  galaxies in  'big gap' groups  is indistinguishable
  with galaxies in other normal groups.}
\label{fig:ms-mh}
\end{figure}

The above  comparison shows  that our  model is  able to  reproduce the
distribution of  the gap seen  in the data, and  now we use  the model
galaxies  to  investigate  the  origin  of  the  'big gap'  groups  in  our
simulation. As to the  TBTF problem in the MW, one  possible solution is
that most  massive subhaloes are  darkness due to their  stochastic star
formation efficiency  (e.g., Guo et  al. 2015).  Here we check  if the
'big gap' groups in our model is  from the stochasticity in star formation.
In Fig.\ref{fig:ms-mh} we  show the stellar mass  versus the accretion
halo mass, $M_{acc}$,  for galaxies in all groups (black  dots) and in
'big gap' groups (red squares).  For satellite galaxies, the accretion halo
mass is the dark matter halo mass  at the time of accretion, and it is
the current halo mass for central galaxy.   It is seen that there is a
good correlation between  halo mass and stellar mass.  The galaxies in
'big gap'  groups have similar distribution  as other  galaxies.
Thus we conclude  that these 'big gap' groups are not  from the stochasticity
of star formation in their satellites.

\begin{figure*}
\centerline{\psfig{figure=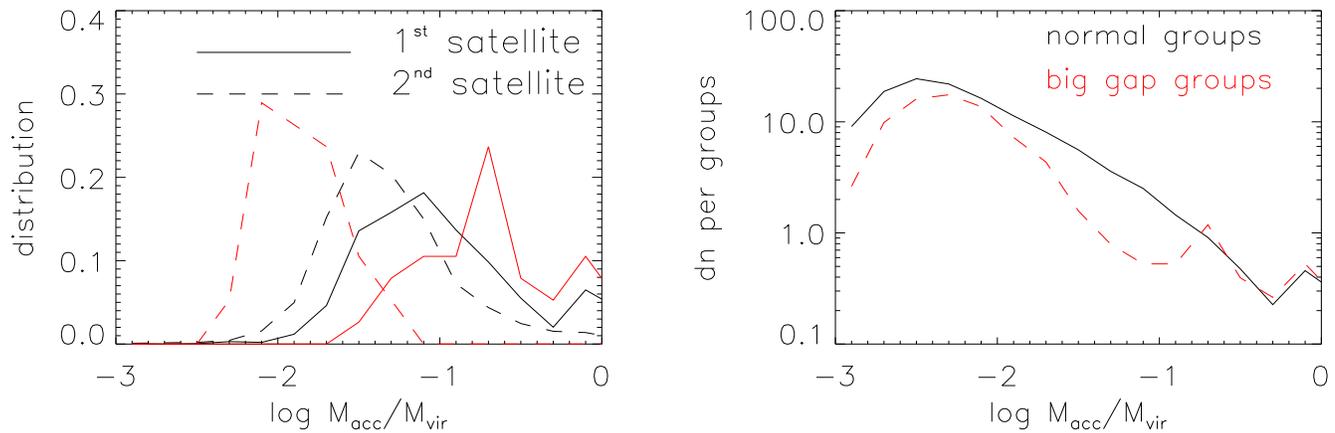,width=0.99\textwidth}}
\caption{The  distribution of  the accretion  halo mass  for satellite
  galaxies. Left pane: the accretion  mass distribution for the first,
  second   massive  satellite   galaxies   using   solid  and   dashed
  lines. Right panel:  the number of accreted haloes as  a function as
  the accretion  mass. In both panel,  red color is for  the 'big gap'
  groups and black for other groups.  A lack of accreted galaxies with
  accretion mass  $M_{acc}/M_{vir} \sim  0.1$ is  clearly seen  in the
  'big gap' groups.}
\label{fig:submf}
\end{figure*}

In Fig.\ref{fig:submf} we show the  distribution of the accretion halo
mass  for  satellite  galaxies  in   groups  with  virial  mass  $\log
M_{vir}=[13,13.5]M_{\odot}$.   The red  lines  are for  the 'big  gap'
groups, and black lines are for other groups.  The left panel show the
distribution  of the  accretion  halo mass  (normalized  by the  group
virial mass) of the most  massive satellite galaxies (solid lines) and
the  second massive  satellites (dashed  lines). It  is found  that in
normal groups the accretion mass  between the first and second massive
satellites are closer,  with the peak mass differ by  about $0.4 dex$.
But for  the 'big gap'  groups, the  second massive satellites  have a
peak  accretion mass  at  about  1\% of  the  group  virial mass.  The
difference between  the first and  second massive is about  $1.5 dex$.
Note that the  peak accretion mass of the first  massive satellites in
'big gap'  groups is slightly higher  than that in the  normal groups.
The right panel  of Fig.~\ref{fig:submf} shows the  halo mass function
of accreted satellite galaxies per  group.  The black solid line shows
that  in  normal  groups,  the  number  of  accreted  low-mass  haloes
increases with  decreasing mass, consistent with  the expectation from
the CDM model.   However, in the 'big  gap' groups, there is  a dip at
around $M_{acc}/M_{vir}  \sim 0.1$ (seen in the left panel),  which happens  to be  between the
first and second massive satellites.   The lack of accreted halos with
$M_{acc}/M_{vir} \sim 0.1$ well explains why there is a big gap in the
stellar mass of satellites in the 'big gap' groups.

\section{Conclusion and Discussion}

Using the  group catalogue  from the SDSS  DR7 and the galaxies
from a semi-analytical model, we study the gap in stellar mass between
the  most  massive  and  the  second  massive  satellite  galaxies  in
groups. We have obtained the following results,

\begin{itemize}

\item The data  and the model have similar distributions  on the gap in
  stellar mass between the most  and second massive satellite galaxies
  in groups. The gap distribution is dependent on group mass and being 
wider in low-mass groups. For  groups with virial mass $\log M_{vir}
  = [13,13.5]M_{\odot}$,  there are  1\% of groups  with a  larger gap
  with $\Delta_{12}>1$ (a  factor of 10) in the data,  and it is about
  2\% in the model.

\item Using model galaxies from the simulation, it is found that there
  is a good  correlation between stellar mass and  accretion halo mass, 
  and the  lower stellar mass in  the second massive  satellites is from
  their lower halo mass at accretion.  The gap in the 'big gap' groups is
  from  the  lack of  accreted  haloes  with  $M_{acc}/M_{vir} \sim  0.1$
  compared  to other normal groups.  The formation  of 'big gap'  groups is
  purely due to their formation history.

\end{itemize}

We note  that in our  analysis we  use the gap  in stellar mass  as an
analog of  the TBTF problem  in the MW.  The often termed TBTF  in the
Milky Way is expressed using  the maximum circular velocity which is a
more reliable  estimator of the  halo mass at accretion.   However, on
groups  scales it  is  expected  that stellar  mass  is more  strongly
correlated  with  the accretion  mass  as  the  stochasticity in  star
formation is  only expected in low-mass haloes.   If the extrapolation
of our  results to the Milky  Way is appropriate, it  implies that the
TBTF problem  in the Milky Way could  be a very nature  outcome of its
formation history although the probability is only around 1\%.

The  lack of  accreted  massive subhaloes  in  the Milky  Way is  also
consistent  with other  expectations.  The  appearance of  two massive
satellites (LMC, SMC) in MW-size  halo is actually in low probability,
and the  accretion of too  many massive subhaloes will  also challenge
the observed stable disk in the Milky Way (e.g., Boylan-Kolchin et al.
2010).   However,  along  with  other  facts on  the  rareness  of  MW
satellite distribution  (great thinner disc and  co-rotated plane), it
is very  likely that the formation  of MW is quite  different from the
typical MW-size halo  in the CDM model. Future  surveys, such as GAIA,
could  put more  strong constraints  on the  formation history  of the
Milky Way.

\section{Acknowledgments}

We thank the anonymous referee for comments to improve the paper. We 
thank Xiaohu Yang for making the group catalogue of SDSS DR7 public
available.   This    work   is   supported   by    the   973   program
(No.   2015CB857003,   2013CB834900),    NSF   of   Jiangsu   Province
(No. BK20140050),  the NSFC (No. 11333008) and  the Strategic Priority
Research Program  the emergence of cosmological structures  of the CAS
(No.  XDB09010403).  The simulations  are  run  on the  Supercomputing
center of PMO.


\begin{thebibliography}{}

\bibitem[]{}
Abazajian K., et al., 2009, ApJS, 182, 543
\bibitem[]{}
Avial-Reese V., Zavala J., Firmani C. \& Hernandez-Toledo H.M., 2008, AJ, 136, 1340
\bibitem[]{}
Bell E., McIntosh D., Katz N \& Weinberg M., 2003, ApJS, 149, 289
\bibitem[]{}
Benson A., Lacey C. G., Baugh C. M., Cole S. \& Frenk C.S., 2002, MNRAS, 333, 156
\bibitem[]{}
Boylan-Kolchin M., Springel V., White S. D. M. \& Jenkins A., 2010, MNRAS, 406, 896
\bibitem[]{}
Boylan-Kolchin M., Bullock J. S. \& Kaplinghat M., 2011, MNRAS, 415, 40
\bibitem[]{}
Boylan-Kolchin M., Bullock J. S. \& Kaplinghat M., 2012, MNRAS, 422, 1203
\bibitem[]{}
Brook C. B. \& Di Cintio A., 2015, MNRAS, 450, 3920
\bibitem[]{}
Brooks A. M., Kuhlen M., Zolotov A. \& Hooper D., 2013, ApJ, 765, 22
\bibitem[]{}
Brooks A. M., \& Zolotov A., 2014, ApJ, 786, 87
\bibitem[]{}
Cautun M, Frenk C. S., van de Weygaert R., Hellwing W. \& Jones B., 2014, MNRAS, 445, 2049
\bibitem[]{}
D'Onghia E., Sommer-Larsen J., Romeo A.D., Burkert A., Pedersen K., Portinari L. \& Rasmussen J., 2005, ApJ, 630, L109 
\bibitem[]{}
Dutton A.,  Macci{\`o} A.V., Frings J., Wang L., Stinson G., Penzo C. \& Kang X., 2016, MNRAS, in press
\bibitem[]{}
Font A. S., et al., 2011, MNRAS, 417, 1260
\bibitem[]{}
Garrison-Kimmel S., Rocha M., Boylan-Kolchin M., Bullock J. S. \& Lally J., 2013, MNRAS, 433, 3539
\bibitem[]{}
Garrison-Kimmel S., Boylan-Kolchin M., Bullock J. S. \& Kirby E., 2014, MNRAS, 444, 222
\bibitem[]{}
Gnedin N. Y., 2000, ApJ, 542, 535
\bibitem[]{}
Governato F., et al., 2012, MNRAS, 422, 1231
\bibitem[]{}
Guo Q. Cooper A., Frenk C. S., Helly J., \& Hellwing W., 2015, MNRAS, 454, 550
\bibitem[]{}
Jiang F. \& van den Bosch F. C., 2015, MNRAS, 453, 3575
\bibitem[]{}
Kang X., Mao S., Gao L., \& Jing Y. P., 2005, A\&A, 437, 383
\bibitem[]{}
Kang X., Li M., Lin W.P. \& Elahi P., 2012, MNRAS, 422, 804
\bibitem[]{}
Kang X., 2014, MNRAS, 437, 3385
\bibitem[]{}
Kennedy R., Frenk C. S., Cole S., \& Benson A., 2014, MNRAS, 442, 2487
\bibitem[]{}
Klypin A., Kravtsov A.V., Valenzuela c., \& Prada F., 1999, MNRAS, 522, 82
\bibitem[]{}
Komatsu E., et al., 2011, ApJS, 192, 18 
\bibitem[]{}
Koposov S., et al., 2008, ApJ, 686, 279
\bibitem[]{}
Kroupa P., Theis C. \& Boily C. M., 2005, A\&A, 431, 517
\bibitem[]{}
Kundert A., et al., 2015, MNRAS, 454, 161
\bibitem[]{}
Libeskind N. I., Frenk C. S., Cole S., Jenkins A. \& Helly J., 2009, MNRAS, 399, 550
\bibitem[]{}
Liu L., Gerke B., Wechsler R., Behroozi P., \& Busha M., 2011, ApJ, 733, 62
\bibitem[]{}
Lovell M., et al., 2012, MNRAS, 420, 2318
\bibitem[]{}
Macci{\`o} A. V., Kang X., Fontanot F., Somerville R. S., Koposov S., \& Monaco P, 2010, MNRAS, 402, 1995
\bibitem[]{}
Macci{\`o} A. V., Ruchayskiy O., Boyarsky A. \& Mu$\tilde{\rm n}$oz-Cuartas J., 2013, MNRAS, 428, 882
\bibitem[]{}
McConnachie A., 2012, AJ, 144, 4
\bibitem[]{}
Moore B., Ghigna S., Governato F., Lake G., Quinn T., Stadel J. \& Tozzi P., 1999, ApJ, 524L, 19
\bibitem[]{}
Pawlowski M. \& Kroupa P., 2013, MNRAS, 435, 2116
\bibitem[]{}
Pawlowski M., Famaey B., Merritt D. \& Kroupa P., 2015, ApJ, 815, 19
\bibitem[]{}
Rocha M., et al., 2013, MNRAS, 430, 81
\bibitem[]{}
Rodriguez-Puebla A., Avila-Reese V. \& Drory N., 2013, ApJ, 773, 172
\bibitem[]{}
Salawa T., et al., 2015, MNRAS, 448, 2941
\bibitem[]{}
Springel V., 2005, MNRAS, 364, 1105
\bibitem[]{}
Strigari L. \& Wechsler R., 2012, ApJ, 749, 75
\bibitem[]{}
Vogelsberger M., Zavala J. \& Loeb A., 2012, MNRAS, 423, 3740
\bibitem[]{}
von Benda-Beckmann A., D'Onghia E., Gottl$\ddot{\rm o}$ber S., Hoeft M., Khalatyan A., Klypin A., M$\ddot{\rm u}$ller V., 2008, MNRAS, 386, 2345
\bibitem[]{}
Wang J.,  Frenk C. S., Navarro J. F., Gao L. \& Sawala T., 2012, MNRAS, 424, 2715
\bibitem[]{}
Wang J., Frenk C. S. \& Cooper A., 2013, MNRAS, 429, 1502
\bibitem[]{}
Yang X., Mo H. J., van den Bosch F. C. \& Jing Y. P., 2005, MNRAS, 356, 1293
\bibitem[]{}
Yang X., Mo H. J., van den Bosch F. C., Zhang Y. \& Han J., 2012, ApJ, 752, 41
\bibitem[]{}
Zentner A., Kravtsov A., Gnedin O. \& Klypin A., 2005, ApJ, 629, 219
\bibitem[]{}
Zolotov A., et al., 2012, ApJ, 761, 71


\end{thebibliography}


\label{lastpage}
\end{document}